\documentclass[10pt, a4paper]{article}
\usepackage[affil-it]{authblk}

\usepackage{citesort}
\usepackage{cite}

\usepackage{cmap}
\usepackage[english]{babel}
\usepackage{amssymb, amsmath}
\usepackage{multicol}
\usepackage{graphicx}
\usepackage{float}
\usepackage[left=2cm,right=2cm,top=2cm,bottom=2cm ,bindingoffset=0cm]{geometry}

\usepackage[usenames]{color}

\begin{document}
\bibliographystyle{abbrv}

\title{Comment on 'Linking Spatial Distributions of Potential and Current in Viscous Electronics'}


\author{M. Semenyakin%
\thanks{\texttt{semenyakinms@gmail.com}}}
\affil{
\small Department of Physics, Taras Shevchenko National University of Kyiv, Kyiv, Ukraine\\
\small Department of Mathematics, NRU HSE, Moscow, Russia\\
\small Skolkovo Institute of Science and Technology, Skolkovo, Russia
}

\maketitle


In \cite{1FalkovichLevitovNature} it was observed that two-dimensional viscous flows of strongly interacting electrons in a confined geometry create vortices, distinguished by the existence of separatrix lines, which separate source-sink stream lines from close loops. In \cite{FalkovichLevitovMagnetic} it was also shown, that properties of these separatrix lines are sensitive to the boundary conditions for the velocities field. Stagnation point's coordinate $x_0$ (point of intersection of separatrix and boundary, where velocity is zero) coincides with the coordinate of the source of current $x_0=0$ in the 'no-slip' case $v_t=0$, and has some distinct coordinate $x_0 = x_*$ in the 'no-stress' case $\partial_n v_t=0$, where $v_t$ - component of speed, tangential to the boundary, and $\partial_n$ - derivative in the direction normal to the boundary.

A class of so-called 'partially slip' boundary conditions is expected to be suitable for 2d viscous electronic flow \cite{1FalkovichLevitovNature}, \cite{GeimExperiment}. It reads: $v_t=l/\eta \,\sigma'_{tn}$, where $\sigma'_{tn}$ - component of viscous stresses tensor and $\eta$ - shear viscosity. In the limit $l\to 0$ it turns into 'no-slip' conditions, while if $l \to \infty$ it becomes 'no-stress', as long as $\sigma_{tn}=\eta(\partial_t v_n+\partial_n v_t)$. Constant $l$ is the 'slip length', characterizing a momentum transfer rate between particles and boundary.

In \cite{GeimExperiment} based on smallness of the Gurzhi effect observed in the experiment with low carrier densities, it was suggested to use the no-stress boundary conditions for computations. In \cite{Molenkamp}, Gurzhi effect was also used to distinguish suitable boundary conditions on the kinetic level. In \cite{WagnerBCS} sufficient efforts was devoted to  predict signatures of boundary conditions, by measuring non-local negative resistance. On the kinetic level, it is unclear yet how to derive hydrodynamic boundary conditions from microscopic one, derived, for instance, in \cite{Soffer}. Here we propose a method based on observation made in \cite{FalkovichLevitovMagnetic}, which will allow to determine a constant $l$ experimentally.

In the regime, considered in \cite{FalkovichLevitovMagnetic}, electrons flow is assumed to have a low Reynolds number \cite{GeimTwoOnOneSide}, \cite{GeimExperiment}, thus it is possible to use Stokes approximation, in which inertia of carriers is neglected:
\begin{equation}
\label{initial}
\eta \Delta v = ne\nabla\phi,
\end{equation}
where $\eta$ - shear viscosity, $ne$ - density of charge, $\phi$ - electro-static potential. As fluid is assumed to be incompressible, we can introduce a stream function $v=e_z \times \nabla\psi$, and by taking a $\mathrm{curl}$ of equation (\ref{initial}) get the bi-harmonic equation. As we have translational invariance along $x$-axis, we can use Fourier transformation in this direction, and finally get:
\begin{equation}
\label{biharmonic}
(\partial_y-k^2)^2 \psi_k(y)=0,\;\; \psi(x,y)=\int\limits_{-\infty}^{+\infty}\psi_k(y) e^{ikx}
\end{equation}
We consider geometry with the source at the point $x=0,y=0$, which is injecting electrons in the strip $w>y>0$. 'Partial slip' boundary conditions for stream function get form $l\partial_{yy}\psi=\partial_y \psi$ at $y=0$, $-l\partial_{yy}\psi_k=\partial_y \psi_k$ at $y=w$. A sign 'minus' in the first condition appears because of the different direction of vector, normal to the boundary. Conditions for the normal component of current $v_y(x,0)=v_y(x,w)=\tilde{I} \delta(x)$ get a form $\psi_k(0)=\psi_k(w)=\tilde{I}/(ik)$. Solving bi-harmonic equation (\ref{biharmonic}), we obtain:
\begin{equation}
\psi_k(y)=\dfrac{\tilde{I}}{i}\dfrac{k (2 l + y) \cosh{k (w - y)} + k (2 l + w - y) \cosh{k y} + (1 + k^2 l y) \sinh{k (w - y)} + (1 + k^2 l (w - y)) \sinh{k y}}{ k (kw +2kl (1 + \cosh{k w}) + \sinh{k w}))}
\end{equation}
In dimensionless quantities $\xi=x/w$, $\epsilon=l/w$, $t=kw$ we get expression for the speed at the edge $y=0$:
\begin{equation}
\label{vfin}
v_x(x,y=0)= \dfrac{\tilde{I}}{\pi w}\int\limits_{0}^{+\infty}\dfrac{t \epsilon (\sinh{t}-t)\sin{t\xi}}{t + \sinh{t} + 2 t \epsilon(1 + \cosh{t})}dt
\end{equation}
To compute profile of speeds, we have to extract part which remains finite at $t \to +\infty$:
\begin{equation}
\label{vFinExtracted}
v_x(x,y=0) = \dfrac{\tilde{I}}{\pi w}\int\limits_{0}^{+\infty}\dfrac{-2t^2 \epsilon (1+\epsilon (1+t+e^{-t}))\sin{(t\xi)}}{(t + \sinh{t} + 2 t \epsilon(1 + \cosh{t}))(1 + 2t\epsilon)}dt + \dfrac{\tilde{I}}{\pi w}\int\limits_{0}^{+\infty}\dfrac{t \epsilon \sin{(t\xi)}}{1 + 2t\epsilon}dt=\dfrac{\tilde{I}}{\pi w}(I_1+I_2)
\end{equation}
Zeroes of this function correspond to the stagnation points of the flow. Numerical computations show the following dependence of the coordinate of stagnation point on the parameter $\epsilon$:
\begin{figure}[H]
\begin{center}
\label{xiofl}
\includegraphics[width=0.5\textwidth]{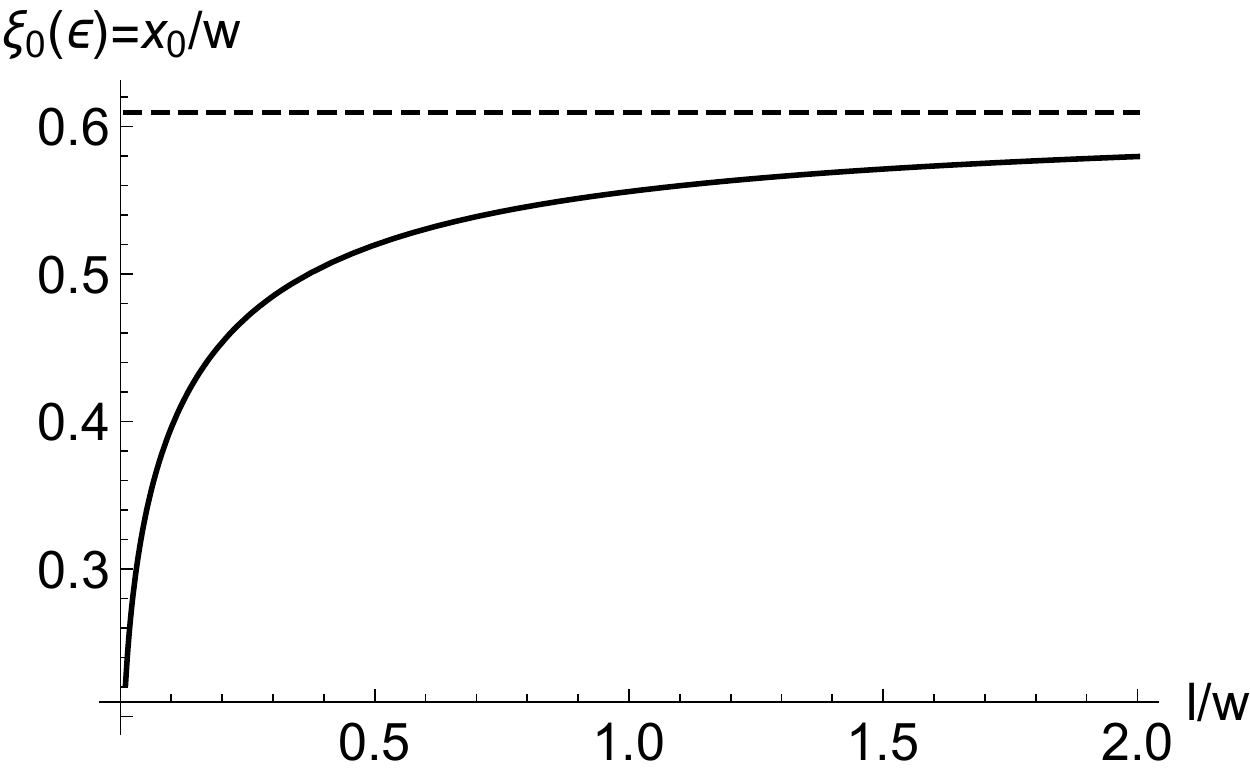}
\caption{Dependence $\xi_0(\epsilon)$ shown by solid line. Dashed line marks asymptotic coordinate $\xi_{\infty}=x/w=0.6095...$ of stagnation point at $\epsilon\to+\infty$}
\end{center}
\end{figure}
We are especially interested in the limit $1\gg\xi\gg\epsilon$, in which power-like scaling at $\epsilon\to0$ for the $\xi_0(\epsilon)$ appears. In this limit we can sufficiently simplify the formula for $v_x$. As expression under integration for $I_1$ decays exponentially at $t \to \infty$, and as pre-factor $t^2\to 0$ at $t\to 0$, main contribution to the integral is given by area $t \sim 1$. In this area we can neglect all terms containing $\epsilon t$ and $\epsilon e^{-t}$ comparing $1$, and as $\xi\ll 1$, we can also assume $\sin(\xi t)\sim \xi t$:
\begin{equation}
I_1 \simeq -2\epsilon \xi \int\limits_{0}^{+\infty}\dfrac{t^3}{t+\sinh{t}}dt=-2A\epsilon \xi
\end{equation}
where $A\simeq 9.93252..\,$. Now, we have to compute $I_2$:
\begin{equation}
I_2=\int\limits_{0}^{+\infty} \dfrac{t \epsilon}{1 + 2t\epsilon}\sin{(t\xi)}dt = \dfrac{\epsilon}{\xi^2}\int\limits_{0}^{+\infty} \dfrac{s\sin{s}}{1 + 2s\lambda}ds
\end{equation}
where $\lambda=\epsilon/\xi=l/x\ll 1$. We can expand it into the asymptotic series:
\begin{equation}
I_2=\dfrac{\epsilon}{\xi^2}\int\limits_{0}^{+\infty} (s-2s^2\lambda+4s^3\lambda^2+...)\sin{(s)}ds
\end{equation}
Defining oscillating integrals as previous, as the principal value, in the first non-vanishing order we get:
\begin{equation}
I_2\simeq 4\lambda\epsilon/\xi^2
\end{equation}
which gives, finally, $(\ref{vFinExtracted})$ in the form
\begin{equation}
v_x \simeq \dfrac{\tilde{I}}{\pi w}(-2A\epsilon \xi+4\lambda\epsilon/\xi^2)
\end{equation}
in the limit $1\gg\xi\gg\epsilon$. Solution of the $v_x=0$ is given by:
\begin{equation}
-2A\epsilon \xi_0+4\lambda_0\epsilon/\xi_0^2=0\;\; \Rightarrow \xi_0=(2\epsilon/A)^{1/4},\;\; x_0^4=2lw^3/A
\end{equation}
which is very close to the results of numerical computations. This result could also be confirmed by dimensional arguments. Indeed, velocity at the boundary consists from two contributions: one dominates near the source $x\ll w$ and coincides with the velocity for half-plane case, $v_{x+} \sim \tilde{I}l^2/x^3$ - lowest power of $x$, which vanishes at $l\to 0$ and is odd with respect to $x\to -x$. Another contribution is related to vortexes and finite width of the stripe, has negative sign, must vanish with $l\to 0$ and be odd under $x\to -x$, too. Thus, we can write $v_{x-}\sim -\tilde{I} lx/w^3$, and condition $v_{x-}+v_{x+}\sim 0$ immediately gives us scaling $x_0^4 \sim l w^3$.\\

Now, we propose the method based on the observation, made in \cite{FalkovichLevitovMagnetic}, which will allow us to measure 'slip'-constant $l$ using weak magnetic field. In magnetic field, normal to the sample's plane, incompressible fluid in Stokes approximation can be described by the equations:
\begin{equation}
-\eta \nabla^2 v_i=-e\partial_i \varphi+\frac{e}{c} [v \times B]_i.
\end{equation}
Following \cite{FalkovichLevitovMagnetic} we take $\mathrm{curl}$ of it
\begin{equation}
\eta\partial\times \Omega=e\partial (\varphi + \frac{B}{c}\psi),
\end{equation}
and using new dynamical variables
\begin{equation}
\label{replacement}
\Omega'=\Omega,\;\;\; \varphi'=\varphi + \frac{B}{c}\psi
\end{equation}
the equations can be rewritten in the initial form, as in the problem without magnetic field. Boundary conditions for the tangential component of velocity remain unchanged, as they are related to the off-diagonal components of momentum flux tensor. Conditions for normal component of velocity are determined by ingoing/outgoing charge, and thus are not affected, too. Stagnation point is characterized by the condition $\partial_y \psi(x_0)=0$. Therefore, we can take derivative of eq. (\ref{replacement}):
\begin{equation}
\partial_y \varphi(B=0)=\partial_y\varphi(B) + \frac{B}{c}\partial_y\psi(B)
\end{equation}
and as $\partial_y \psi(x_0)=0$ at the stagnation point, and $\partial_x \psi(x_0)=0$ on the whole boundary, except source, get:
\begin{equation}
\vec{E}(B=0,x_0)=\vec{E}(B,x_0)
\end{equation}
That means that the electric field at the stagnation point doesn't depend on the external magnetic field - in contrast to all other points, as far as speed on the boundary is not vanishing there. But we have predicted theoretically dependence $x_0=x_0(l/w)$ of the coordinate of stagnation point on the $l$. So, if we will measure a position of stagnation point, we would know $l$.

\section*{Acknowlegments}

This work was done during Kupcinet-Getz International Summer School at Weizmann Institute. Author is highly grateful to Gregory Falkovich, who introduced him to the field of study and guided through the research. He has also suggested dimensional reasoning for the stagnation point's asymptotic. Author is also appreciated to Pavlo Gavrylenko for fruitful discussions. 

\bibliography{MyCitations}

\end{document}